# New Lithium- and Diamines-Intercalated Superconductors Li$_x$(C$_2$H$_8$N$_2$)$_y$TiSe$_2$ and Li$_x$(C$_6$H$_{16}$N$_2$)$_y$TiSe$_2$

Kazuki Sato, Takashi Noji, Takehiro Hatakeda, Takayuki Kawamata,
Masatsune Kato, and Yoji Koike

*Department of Applied Physics, Tohoku University, 6-6-05 Aoba, Aramaki, Aoba-ku,
Sendai 980-8579, Japan*

New intercalation superconductors of Li$_x$(C$_2$H$_8$N$_2$)$_y$TiSe$_2$ and Li$_x$(C$_6$H$_{16}$N$_2$)$_y$TiSe$_2$ with $T_c$ = 4.2 K have successfully been synthesized via the co-intercalation of lithium and ethylenediamine or hexamethylenediamine into 1T-TiSe$_2$. Moreover, it has been found that both intercalation compounds of Li$_x$TiSe$_2$ and (C$_2$H$_8$N$_2$)$_y$TiSe$_2$ also show superconductivity with $T_c$ = 2.4 K and 2.8 K, respectively. These results indicate that both the electron doping due to the intercalation of lithium and the expansion of the interlayer spacing between TiSe$_2$ layers due to the intercalation of diamines suppress the charge density wave in 1T-TiSe$_2$, leading to the appearance of superconductivity.

## 1. Introduction

Some of transition metal dichalcogenides (TMDs) $MX_2$ ($M$ = transition metal, $X$ = S, Se, or Te) with layered structures are famous for intriguing phenomena such as anisotropic superconductivity and charge density wave (CDW), owing to their quasi-two-dimensional electronic structures.[1,2] Among TMDs, 1T-TiSe$_2$ shown in Fig. 1(a) has attracted particular attention, because it undergoes a CDW phase transition at ~ 220 K in spite of the semimetallic electronic structure.[3] The CDW in 1T-TiSe$_2$ is characterized by the three-dimensional commensurate wave vector (2 × 2 × 2) connecting the Γ ($k_z$ = 0) and L points ($k_z$ = π/$c$, where $c$ is the $c$-axis length) in the hexagonal Brillouin zone. Early studies have suggested that the CDW arises from the conventional Fermi-surface nesting between the hole pocket due to Se 4$p$ electrons at the Γ point and the electron pocket due to Ti 3$d$ electrons at the L point.[3,4] Recent studies, however, have proposed various scenarios on the origin of CDW such as band Jahn-Teller mechanisms[5-10] and excitonic condensation mechanisms,[11-18] based on the consideration that both the small pockets at Γ and L points are spherical and therefore unfavorable for the nesting. Although 1T-TiSe$_2$ shows no superconductivity, it has recently been reported that superconductivity with the superconducting transition temperature $T_c$ = 2 – 4

K is induced by the suppression of the CDW transition in 1T-TiSe$_2$ through the intercalation of Cu or Pd,[19-26] the elemental substitution[27] and the application of high pressure[28,29] or electric field.[30] These studies indicate that the superconductivity appears around the quantum critical point at which the CDW transition temperature $T_{CDW}$ goes to 0 K, as in the case of some other superconductors where superconductivity appears around a quantum critical point.[31,32]

In this paper, we report on the successful synthesis of new intercalation superconductors of Li$_x$(C$_2$H$_8$N$_2$)$_y$TiSe$_2$ and Li$_x$(C$_6$H$_{16}$N$_2$)$_y$TiSe$_2$ with $T_c$ = 4.2 K via the co-intercalation of lithium and ethylenediamine (EDA), C$_2$H$_8$N$_2$, or hexamethylenediamine (HMDA), C$_6$H$_{16}$N$_2$, into 1T-TiSe$_2$. We also report on the synthesis of new intercalation superconductors of Li$_x$TiSe$_2$ with $T_c$ = 2.4 K and (C$_2$H$_8$N$_2$)$_y$TiSe$_2$ with $T_c$ = 2.8 K. Moreover, effects of the intercalation and deintercalation on the CDW in 1T-TiSe$_2$ have also been investigated.

## 2. Experimental

Polycrystalline host samples of 1T-TiSe$_2$ were prepared by the solid-state reaction method. Titanium powder and selenium grains were weighted in a molar ratio of Ti : Se = 1 : 2, mixed thoroughly and pressed into pellets. The pellets were sealed in an evacuated quartz tube and heated at 800°C for 150 h. The obtained pellets of 1T-TiSe$_2$ were pulverized into powder to be used for the intercalation. Only Li-intercalated samples of Li$_x$TiSe$_2$ were prepared using 1.6 M solution of $n$-butyllithium dissolved in hexane. The powdery 1T-TiSe$_2$ was put into the solution in a molar ratio of Li : 1T-TiSe$_2$ = 1 : 1 and reacted at room temperature for 60 h in air. Both Li- and EDA- or HMDA-intercalated samples of Li$_x$(C$_2$H$_8$N$_2$)$_y$TiSe$_2$ or Li$_x$(C$_6$H$_{16}$N$_2$)$_y$TiSe$_2$ were prepared as follows.[33,34] An appropriate amount of the powdery 1T-TiSe$_2$ was placed in a beaker filled with 0.01 M solution of pure Li metal dissolved in EDA or HMDA. The amount of 1T-TiSe$_2$ was calculated in a molar ratio of Li : 1T-TiSe$_2$ = 0 – 1 : 1. The reaction was carried out at 50°C for 2 weeks. The product in residual EDA was washed with fresh EDA. On the other hand, the separation of the product in residual HMDA was easily performed by the solidification of residual HMDA at the top cap of the beaker, keeping the temperature of the top cap of the beaker below the melting point of HMDA (42°C). All the processes were carried out in an argon-filled glove box. For the investigation of the change of physical properties by the deintercalation of EDA, the as-intercalated samples were placed in a glass tube evacuated using a rotary pump and annealed at 250°C for 20 h. Here, it is noted that $x$ values in intercalated samples of Li$_x$TiSe$_2$, Li$_x$(C$_2$H$_8$N$_2$)$_y$TiSe$_2$ and Li$_x$(C$_6$H$_{16}$N$_2$)$_y$TiSe$_2$ are nominal ones based on the assumption that all Li$^+$ ions in the solution are intercalated into 1T-TiSe$_2$.[34]

Both the host sample of 1T-TiSe$_2$ and intercalated samples were characterized by the powder x-ray diffraction using CuK$_\alpha$ radiation.  For the intercalated samples, an airtight sample-holder was used.  The diffraction patterns were analyzed using RIETAN-FP.[35]  The magnetic susceptibility $\chi$ was measured using a superconducting quantum interference device (SQUID) magnetometer (Quantum Design, MPMS).  Measurements of the electrical resistivity $\rho$ were also carried out by the standard dc four-probe method.  For the $\rho$ measurements, both the powdery host sample of 1T-TiSe$_2$ and powdery as-intercalated/deintercalated samples were pressed into pellets at room temperature without heat treatment.  Thermogravimetric (TG) measurements were performed in flowing gas of argon, using a commercial analyzer (SII Nano Technology Inc., TG/DTA7300).

## 3. Results and Discussion

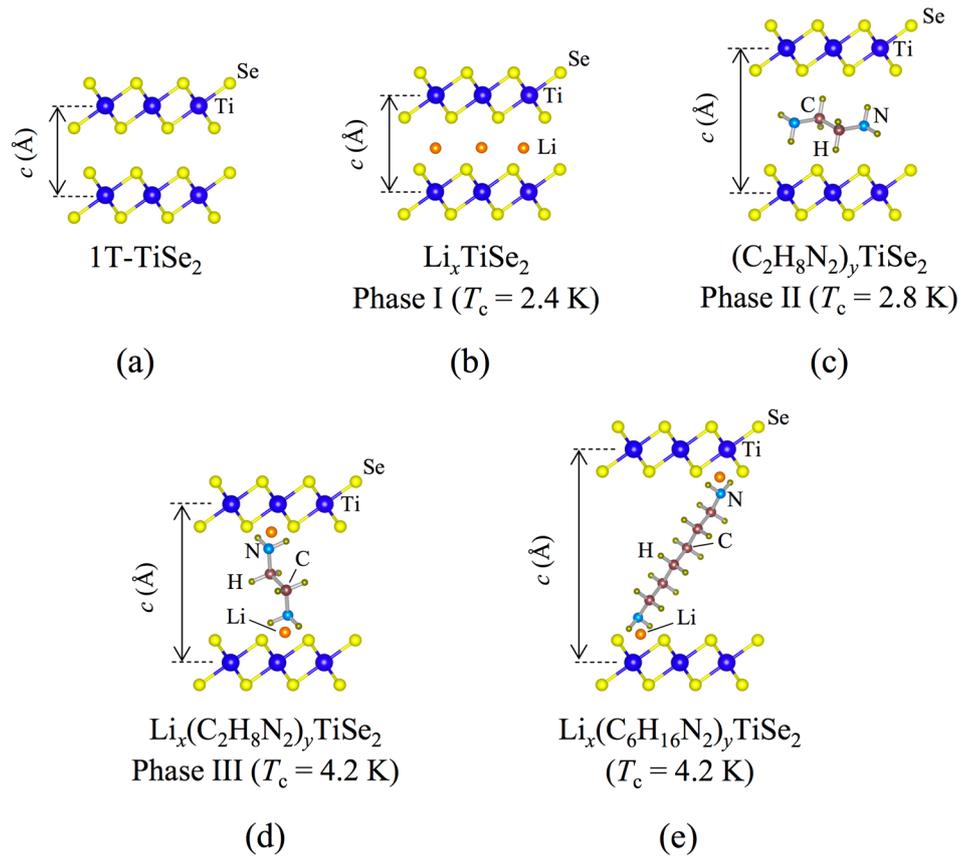

Fig. 1. Schematic views of crystal structures of (a) 1T-TiSe$_2$, (b) Li$_x$TiSe$_2$, (c) (C$_2$H$_8$N$_2$)$_y$TiSe$_2$, (d) Li$_x$(C$_2$H$_8$N$_2$)$_y$TiSe$_2$, (e) Li$_x$(C$_6$H$_{16}$N$_2$)$_y$TiSe$_2$.

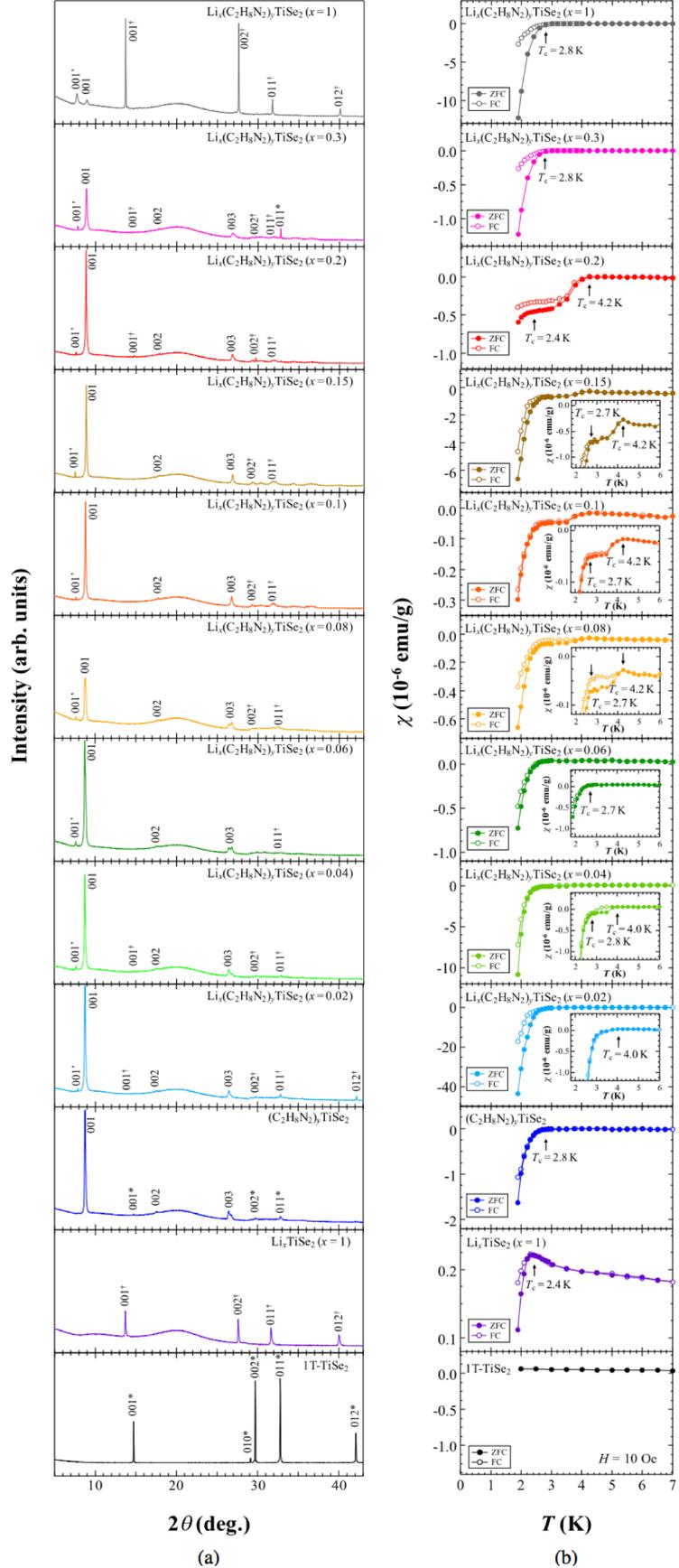

Fig. 2. (a) Powder x-ray diffraction patterns of the host compound of 1T-TiSe$_2$ and intercalated samples of Li$_x$TiSe$_2$ ($x$ = 1), (C$_2$H$_8$N$_2$)$_y$TiSe$_2$ and Li$_x$(C$_2$H$_8$N$_2$)$_y$TiSe$_2$ ($x$ = 0.02 − 1), using CuKα radiation. Indices with an asterisk are due to 1T-TiSe$_2$. Indices with a dagger, no mark and a quotation mark represent Phase I, Phase II and Phase III of Li$_x$(C$_2$H$_8$N$_2$)$_y$TiSe$_2$, respectively. All indices are based on the 1T-type structure ($P\bar{3}m1$). The broad peak around $2\theta$ = 20° is due to the airtight sample-holder. (b) Temperature dependences of the magnetic susceptibility $\chi$ in a magnetic field of 10 Oe on zero-field cooling (ZFC) (closed circles) and on field cooling (FC) (open circles) for the host sample of 1T-TiSe$_2$ and intercalated samples of Li$_x$TiSe$_2$ ($x$ = 1), (C$_2$H$_8$N$_2$)$_y$TiSe$_2$ and Li$_x$(C$_2$H$_8$N$_2$)$_y$TiSe$_2$ ($x$ = 0.02 − 1).

Figure 2(a) shows the powder x-ray diffraction patterns of the host sample of 1T-TiSe$_2$ and intercalated samples of Li$_x$TiSe$_2$ ($x$ = 1), (C$_2$H$_8$N$_2$)$_y$TiSe$_2$ and Li$_x$(C$_2$H$_8$N$_2$)$_y$TiSe$_2$ ($x$ = 0.02 – 1). The broad peak around $2\theta$ = 20˚ is due to the airtight sample-holder. The powder x-ray diffraction patterns of pristine 1T-TiSe$_2$ and only Li-intercalated Li$_x$TiSe$_2$ ($x$ = 1) are well indexed based on the 1T-type structure ($P\bar{3}m1$). The lattice parameters are estimated to be $a$ = 3.536(1) Å, $c$ = 6.021(5) Å for 1T-TiSe$_2$ and $a$ = 3.64(3) Å, $c$ = 6.45(7) Å for Li$_x$TiSe$_2$, which are consistent with those in the previous reports.[3,36] The present results of Li$_x$TiSe$_2$ indicate that Li$^+$ ions are intercalated into the octahedral site in the van der Waals gap between TiSe$_2$ layers, as shown in Fig. 1(b). Here, we call this phase Phase I of Li$_x$(C$_2$H$_8$N$_2$)$_y$TiSe$_2$ with $y$ = 0. For only EDA-intercalated (C$_2$H$_8$N$_2$)$_y$TiSe$_2$, new Bragg peaks are observed at $2\theta$ ~ 9˚ and ~ 26˚, though there still remain small Bragg peaks due to non-intercalated regions of 1T-TiSe$_2$. The $c$-axis length of (C$_2$H$_8$N$_2$)$_y$TiSe$_2$ is estimated to be 10.107(4) Å, so that the $c$-axis is expanded by ~ 4.1 Å through the intercalation of EDA. Since the width and length of EDA are approximately 3.7 Å and 5.1 Å, respectively, it seems that EDA is intercalated in parallel between TiSe$_2$ layers, as shown in Fig. 1(c). Here, we call this phase Phase II of Li$_x$(C$_2$H$_8$N$_2$)$_y$TiSe$_2$ with $x$ = 0. As for both Li- and EDA-intercalated samples of Li$_x$(C$_2$H$_8$N$_2$)$_y$TiSe$_2$ ($x$ = 0.02 – 1), new Bragg peaks are observed at $2\theta$ ~ 8˚, 9˚ and ~ 27˚, though there still remain regions where EDA is not intercalated in a sample. The new Bragg peaks are regarded as being due to Phase II and a different phase (we call this phase Phase III) with the $c$-axis lengths of ~ 10 Å and ~ 11.6 Å, respectively, of the intercalation compound of Li$_x$(C$_2$H$_8$N$_2$)$_y$TiSe$_2$, as listed in Table I. Phase II is the same as that observed in (C$_2$H$_8$N$_2$)$_y$TiSe$_2$, though Li$^+$ ions may be included dilutely in Phase II of both Li- and EDA-intercalated samples. In the case of Phase III, on the other hand, the expansion of the $c$-axis is ~ 5.6 Å, so that EDA is speculated to be intercalated perpendicularly to TiSe$_2$ layers, as shown in Fig. 1(d). This crystal structure is analogous to those of both Li- and EDA-intercalated Li$_x$(C$_2$H$_8$N$_2$)$_y$Fe$_{2-z}$Se$_2$. That is, Li$^+$ ions are located near Se atoms and lone-pair electrons of N atoms in EDA are attracted by Li$^+$ ions. [37,38] In both Li- and EDA-intercalated Li$_x$(C$_2$H$_8$N$_2$)$_y$TiSe$_2$ with $x$ = 1, Phase I of Li$_x$TiSe$_2$ is dominant, though the reason is not clear.

Figure 2(b) shows the temperature dependences of $\chi$ in a magnetic field of 10 Oe on zero-field cooling (ZFC) and on field cooling (FC) for the host sample of 1T-TiSe$_2$ and intercalated samples of Li$_x$TiSe$_2$ ($x$ = 1), (C$_2$H$_8$N$_2$)$_y$TiSe$_2$ and Li$_x$(C$_2$H$_8$N$_2$)$_y$TiSe$_2$ ($x$ = 0.02 – 1). Although pristine 1T-TiSe$_2$ shows no superconductivity, a superconducting transition is observed at 2.4 K in only Li-intercalated Li$_x$TiSe$_2$ ($x$ = 1). Large positive values of $\chi$ in the normal state may be due to magnetic impurities taken into the sample in the intercalation process. It is found that only EDA-intercalated $(C_2H_8N_2)_y TiSe_2$ also shows superconductivity with $T_c = 2.8$ K.

Table I. $c$-axis lengths and $T_c$'s of the host compound 1T-TiSe$_2$, as-intercalated samples and post-annealed (250˚C, 20 h in vacuum) samples.  Here, the Li content $x$ is the nominal one.  $T_c$ values obtained in the magnetic susceptibility ($\chi$ - $T$) and electrical resistivity ($\rho$ - $T$) measurements are defined at onset temperatures where $\chi$ and $\rho$ start to deviate from the normal-state values.  For bracketed $T_c$ values obtained in the $\rho$ - $T$ measurements, the relevant phase is not clear.

| | Li content ($x$) | $c$ (Å) Phase I | $c$ (Å) Phase II | $c$ (Å) Phase III | $T_c$ (K) ($\chi$ - $T$) | $T_c$ (K) ($\rho$ - $T$) |
|---|---|---|---|---|---|---|
| 1T-TiSe$_2$ (host) | 0 | 6.021(5) | | | - | - |
| Li$_x$TiSe$_2$ | 1 | 6.45(7) | | | 2.4 | |
| Li$_x$(C$_2$H$_8$N$_2$)$_y$TiSe$_2$ | 0 | | 10.107(4) | | 2.8 | 2.15 |
| | 0.02 | 6.02(1) | 10.10(1) | 11.44(1) | 4.0 | 2.73 |
| | 0.04 | 6.03(8) | 10.117(5) | 11.57(1) | 2.8 / 4.0 | (2.42) |
| | 0.06 | 6.043(8) | 10.12(1) | 11.62(5) | 2.7 | 1.51 |
| | 0.08 | 6.07(6) | 10.10(8) | 11.581(5) | 2.7 / 4.2 | (1.38) |
| | 0.1 | 6.08(9) | 9.97(8) | 11.53(3) | 2.7 / 4.2 | (0.97) |
| | 0.15 | 6.09(7) | 9.94(9) | 11.59(5) | 2.7 / 4.2 | (0.78) |
| | 0.2 | 6.11(8) | 9.96(7) | 11.55(2) | 2.4 / 4.2 | (0.84) |
| | 0.3 | 6.13(1) | 9.96(3) | 11.52(1) | 2.8 | 0.7 |
| | 1 | 6.45(7) | 10.12(1) | 11.43(7) | 2.8 | - |
| Li$_x$(C$_6$H$_{16}$N$_2$)$_y$TiSe$_2$ | 0.1 | 6.02(1) | 14.07(1) | | 4.2 | 2.57 |
| | 0.2 | 6.02(2) | 14.08(9) | | 4.2 | 2.66 |
| (C$_2$H$_8$N$_2$)$_y$TiSe$_2$ (post-annealed) | 0 | 6.0(6) | | | - | - |
| Li$_x$(C$_2$H$_8$N$_2$)$_y$TiSe$_2$ (post-annealed) | 0.2 | 6.13(8) | | | 2.4 | - |

Here, $T_c$ is defined at the onset temperature where $\chi$ starts to deviate from the normal-state value. As for both Li- and EDA-intercalated samples of $Li_x(C_2H_8N_2)_yTiSe_2$ ($x = 0.02 – 1$), two steps of superconducting transition are observed: the first and second transitions are at 4.0 – 4.2 K and 2.4 – 2.8 K, respectively. Taking into account the powder x-ray diffraction results and $T_c = 2.8$ K of $(C_2H_8N_2)_yTiSe_2$, the second at 2.4 – 2.8 K is regarded as being mainly due to Phase II of $Li_x(C_2H_8N_2)_yTiSe_2$, and therefore the first at 4.0 – 4.2 K is inferred to be due to Phase III of $Li_x(C_2H_8N_2)_yTiSe_2$, as listed in Table I. The $T_c$ value of Phase III of $Li_x(C_2H_8N_2)_yTiSe_2$ ($x = 0.02 – 0.2$) is not so dependent on $x$. This is inferred to be because distribution of $Li^+$ ions in a sample is so inhomogeneous that there remain regions where $Li^+$ ions are not intercalated and therefore the real $x$ value in Li- and EDA-intercalated regions is not so dependent on the nominal $x$ one. Accordingly, the real $x$ value in Li- and EDA-intercalated regions is estimated to be larger than the nominal $x$ value. In both Li- and EDA-intercalated $Li_x(C_2H_8N_2)_yTiSe_2$ with $x = 0.3$, no superconducting transition due to Phase III is observed. This may be caused by the overdoping of electrons due to the intercalation of a large amount of $Li^+$ ions. In addition, it is noted that superconducting volume fractions of these samples estimated from the change of $\chi$ below $T_c$ are as small as less than 1 %. One reason is that $\chi$ does not decrease completely at the measured lowest temperature of 2 K. The others may be the inclusion of non-superconducting non-intercalated regions of 1T-TiSe$_2$ in a sample and the use of the powdery samples. In the case of both Na- and NH$_3$-intercalated $Na_x(NH_3)_yMoSe_2$, in fact, it has been reported that the superconducting volume fraction estimated from $\chi$ measurements is much smaller in the polycrystalline powdery sample than in the single crystal.[39]

Figure 3(a) shows the powder x-ray diffraction patterns of the host sample of 1T-TiSe$_2$ and both Li- and HMDA-intercalated samples of $Li_x(C_6H_{16}N_2)_yTiSe_2$ ($x = 0.1, 0.2$). New Bragg peaks are observed owing to the intercalation, though there remain Bragg peaks due to non-intercalated regions of 1T-TiSe$_2$. The powder x-ray diffraction patterns of both Li- and HMDA-intercalated $Li_x(C_6H_{16}N_2)_yTiSe_2$ ($x = 0.1, 0.2$) are well indexed based on the 1T-type structure ($P\bar{3}m1$), and the $c$-axis length is estimated to be 14.07(1) Å and 14.08(9) Å for $x = 0.1$ and 0.2, respectively. As listed in Table I, these $c$-axis lengths are much large than those of both Li- and EDA-intercalated samples, and the $c$-axis is expanded by ~ 8.1 Å through the co-intercalation of Li and HMDA. Since the width and length of HMDA are approximately 3.7 Å and 10.4 Å, respectively, it is speculated that HMDA is obliquely intercalated between TiSe$_2$ layers, as shown in Fig. 1(e). Such an obliquely intercalated structure is speculated to exist in both Li- and HMDA-intercalated $Li_x(C_6H_{16}N_2)_yFe_{2-z}Se_2$ also.[38,40]

The temperature dependences of $\chi$ in a magnetic field of 10 Oe on ZFC and FC for both Li- and HMDA-intercalated $Li_x(C_6H_{16}N_2)_yTiSe_2$ ($x = 0.1, 0.2$) are shown in Fig. 3(b).

Both samples show a superconducting transition at 4.2 K. Since superconducting volume fractions of these samples estimated from the change of $\chi$ below $T_c$ are small, the real $x$ in intercalated regions of $Li_x(C_6H_{16}N_2)_yTiSe_2$ is estimated to be much larger than the nominal $x$ value, as in the case of $Li_x(C_2H_8N_2)_yTiSe_2$. The $T_c$ value is nearly the same as that of Phase III of both Li- and EDA-intercalated $Li_x(C_2H_8N_2)_yTiSe_2$, though the interlayer distance is expanded more than that of Phase III of $Li_x(C_2H_8N_2)_yTiSe_2$. This suggests that the electronic state is nearly the same between these samples.

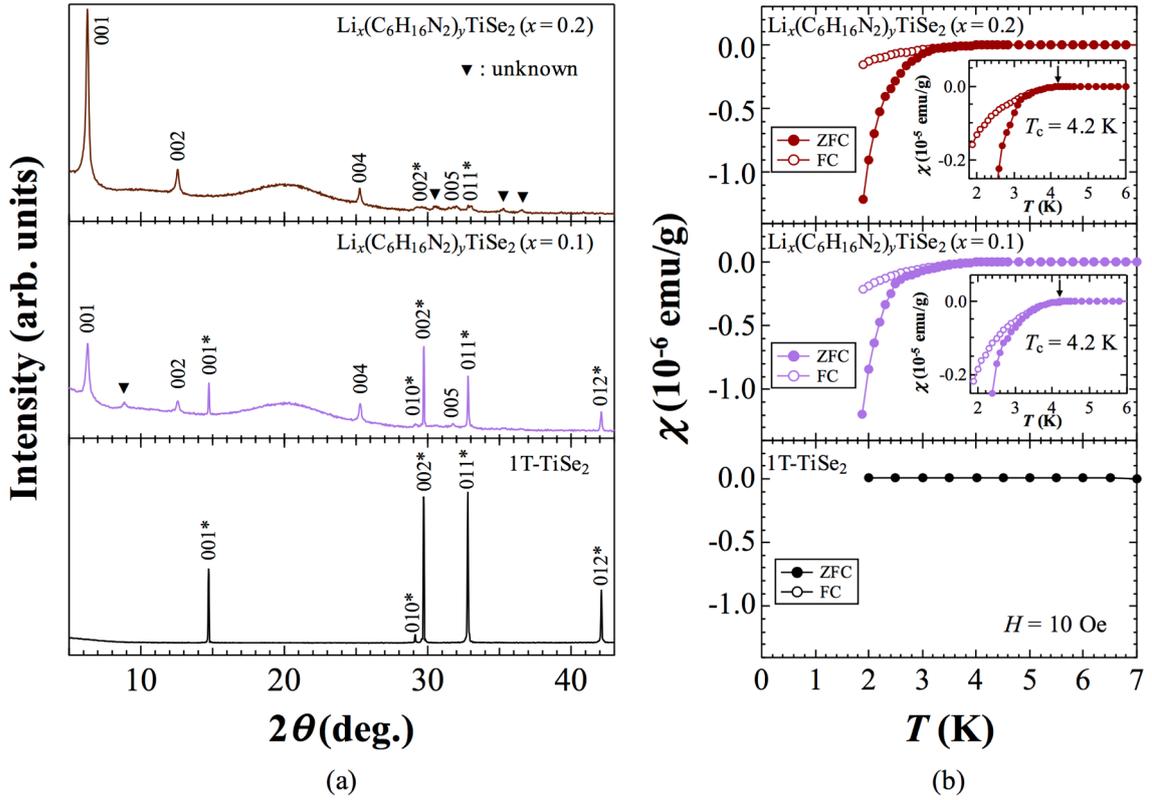

Fig. 3. (a) Powder x-ray diffraction patterns of the host sample of 1T-$TiSe_2$ and both Li- and HMDA-intercalated samples of $Li_x(C_6H_{16}N_2)_yTiSe_2$ ($x$ = 0.1, 0.2), using CuKα radiation. Indices without and with an asterisk are due to $Li_x(C_6H_{16}N_2)_yTiSe_2$ and 1T-$TiSe_2$, respectively. Peaks marked by ▼ are due to an unknown compound. All indices are based on the 1T-type structure ($P\bar{3}m1$). The broad peak around $2\theta$ = 20° is due to the airtight sample-holder. (b) Temperature dependences of the magnetic susceptibility $\chi$ in a magnetic field of 10 Oe on zero-field cooling (ZFC) (closed circles) and on field cooling (FC) (open circles) for the host sample of 1T-$TiSe_2$ and both Li- and HMDA-intercalated samples of $Li_x(C_6H_{16}N_2)_yTiSe_2$ ($x$ = 0.1, 0.2).

The superconductivity observed in the χ measurements for the intercalated samples has been confirmed in the ρ measurements as follows. Figures 4(a) and (b) show the temperature dependences of ρ for the host sample of 1T-TiSe$_2$ and intercalated samples of Li$_x$(C$_2$H$_8$N$_2$)$_y$TiSe$_2$ ($x$ = 0 − 1) and Li$_x$(C$_6$H$_{16}$N$_2$)$_y$TiSe$_2$ ($x$ = 0.1, 0.2). It is found that 1T-TiSe$_2$ exhibits a behavior of ρ well known in single-crystalline samples.[28] That is, with decreasing temperature, ρ increases a little at high temperatures, increases dramatically below $T_{CDW}$ ~ 220 K, shows a broad hump around 150 K and decreases at low temperatures. The upturn behavior of ρ at very low temperatures is due to insulating grain-boundaries in the pelletized samples, because no upturn has been observed in single crystalline samples of 1T-TiSe$_2$.

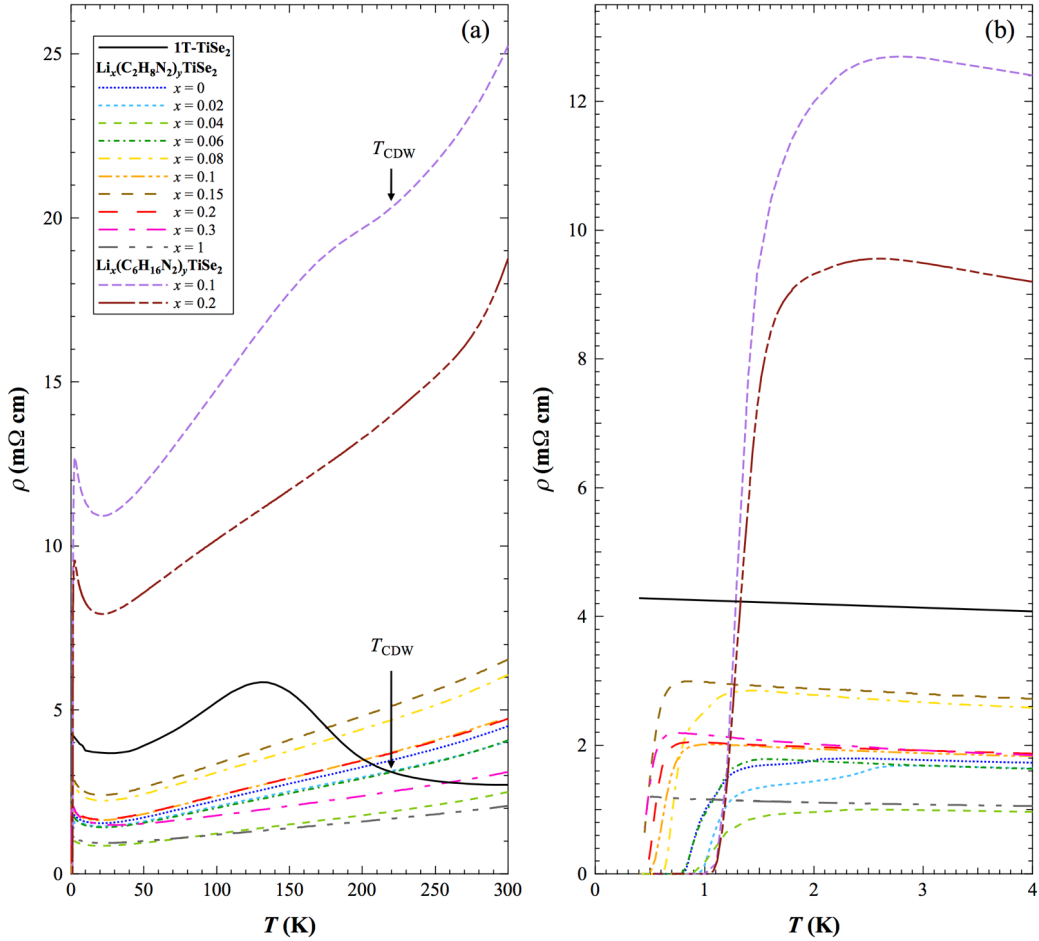

Fig. 4. (a) Temperature dependences of the electrical resistivity ρ for the host sample of 1T-TiSe$_2$ and intercalated samples of Li$_x$(C$_2$H$_8$N$_2$)$_y$TiSe$_2$ ($x$ = 0 − 1) and Li$_x$(C$_6$H$_{16}$N$_2$)$_y$TiSe$_2$ ($x$ = 0.1, 0.2), which were pelletized at room temperature without heat treatment. (b) Enlarged view of Fig. (a) at low temperatures around $T_c$.

As for intercalated samples of $Li_x(C_2H_8N_2)_yTiSe_2$ ($x = 0 - 1$) and $Li_x(C_6H_{16}N_2)_yTiSe_2$ ($x = 0.1$, 0.2), $\rho$ shows a metallic temperature-dependence except for the upturn of $\rho$ at very low temperatures due to insulating grain-boundaries.  It is found that the CDW transition is completely suppressed, though there slightly remains the transition for $Li_x(C_6H_{16}N_2)_yTiSe_2$ with $x = 0.1$.  As shown in Fig. 4(b), all the intercalated samples show a superconducting transition at a low temperature below 3 K.  Values of $T_c$, defined at the onset temperature where $\rho$ starts to deviate from the normal-state value, are listed in Table I.  These results indicate that both the intercalation of EDA and the co-intercalation of Li and EDA or HMDA suppress the CDW transition and induce a metallic state, leading to the appearance of superconductivity.  The coexistence of the CDW transition and superconductivity in $Li_x(C_6H_{16}N_2)_yTiSe_2$ with $x = 0.1$ is due to the phase separation of non-intercalated regions of 1T-TiSe$_2$ undergoing the CDW transition and Li- and HMDA-intercalated regions undergoing the superconductivity, taking into account the powder x-ray diffraction patterns shown in Fig. 3(a).  Here, it is noted that the $T_c$ values obtained in the $\rho$ measurements are lower than those obtained in the $\chi$ measurements, as listed in Table I.  This is probably due to the degradation of the intercalated samples caused by the atmospheric exposure in the process of making four terminals on the sample surface for the $\rho$ measurements.

Figure 5 shows the temperature dependences of $\chi$ in a magnetic field of 1 T on ZFC for the host sample of 1T-TiSe$_2$ and intercalated samples of $Li_xTiSe_2$ ($x = 1$) and $Li_x(C_2H_8N_2)_yTiSe_2$ ($x = 0 - 1$).  In 1T-TiSe$_2$, $\chi$ is found to decrease at low temperatures below $T_{CDW} \sim 220$ K, which is due to the decrease in the Pauli paramagnetism caused by the opening of the CDW gap.[3]  For the intercalated samples of $Li_xTiSe_2$ ($x = 1$) and $Li_x(C_2H_8N_2)_yTiSe_2$ ($x = 0 - 1$), on the other hand, no decrease in $\chi$ is observed, indicating that the CDW transition is completely suppressed as confirmed in the $\rho$ results shown in Fig. 4(a).  It is found that $\chi$ tends to increase with increasing $x$.  This suggests that intercalated Li supplies TiSe$_2$ layers with electron carriers, leading to the increase in the Pauli paramagnetism.  It is also found that $\chi$ of $(C_2H_8N_2)_yTiSe_2$ with $x = 0$ is larger than that of 1T-TiSe$_2$, which may be due to the increase in the electronic density of states at the Fermi level owing to the enhancement of the two-dimensionality of the electronic structure caused by the expansion of the interlayer spacing between TiSe$_2$ layers.

For the deintercalation of EDA, TG measurements were carried out.  Figure 6 shows TG curves on heating up to 900°C at a rate of 1°C/min for the host sample of 1T-TiSe$_2$ and intercalated samples of $(C_2H_8N_2)_yTiSe_2$ and $Li_x(C_2H_8N_2)_yTiSe_2$ ($x = 0.1$, 0.2).  For the intercalated samples, several steps of mass loss are observed.  These steps are similar to those observed for both Li- and EDA-intercalated $Li_x(C_2H_8N_2)_yFe_{2-z}Se_2$.[41]  Accordingly, it is inferred that the first loss below $\sim 130$ °C is due to the desorption of EDA on the surface of grains and that the second loss between ~ 130 °C and 200 °C is due to the deintercalation of EDA. The EDA contents $y$ in the as-intercalated samples of Li$_x$(C$_2$H$_8$N$_2$)$_y$TiSe$_2$ ($x$ = 0, 0.1, 0.2) are estimated from the second loss to be ~ 0.3.

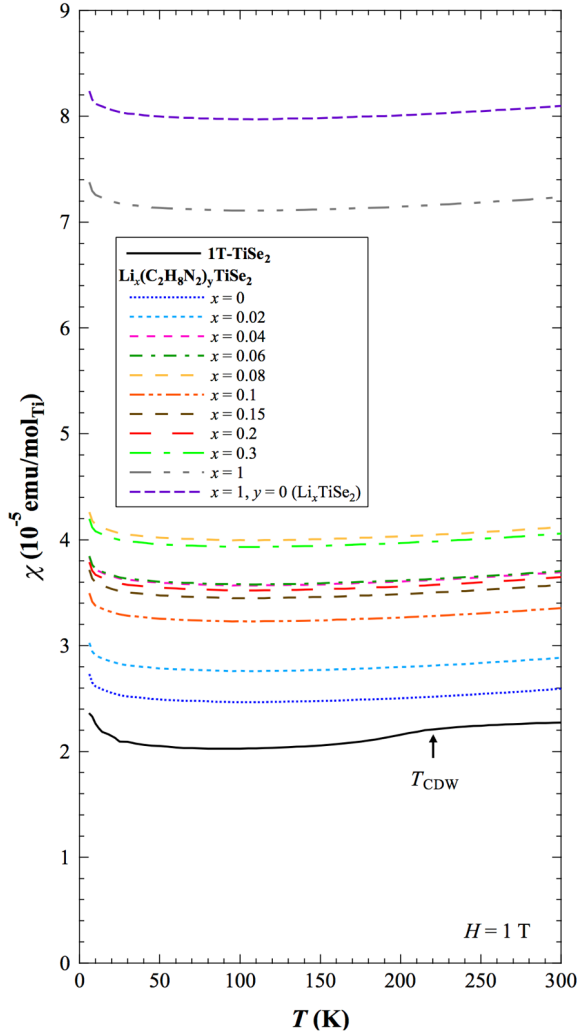

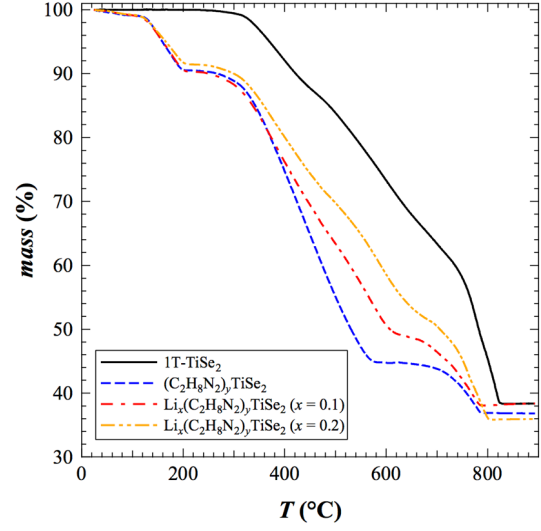

Fig. 6. Thermogravimetric curves on heating at a rate of 1 °C/min for the host sample of 1T-TiSe$_2$ and intercalated samples of (C$_2$H$_8$N$_2$)$_y$TiSe$_2$ and Li$_x$(C$_2$H$_8$N$_2$)$_y$TiSe$_2$ ($x$ = 0.1, 0.2).

Fig. 5. Temperature dependences of the magnetic susceptibility $\chi$ in a magnetic field of 1 T on zero-field cooling for the host sample of 1T-TiSe$_2$ and intercalated samples of Li$_x$TiSe$_2$ ($x$ = 1), (C$_2$H$_8$N$_2$)$_y$TiSe$_2$ and Li$_x$(C$_2$H$_8$N$_2$)$_y$TiSe$_2$ ($x$ = 0.02 – 1).

Figure 7(a) shows the powder x-ray diffraction patterns of (C$_2$H$_8$N$_2$)$_y$TiSe$_2$ and Li$_x$(C$_2$H$_8$N$_2$)$_y$TiSe$_2$ ($x$ = 0.2) samples post-annealed in vacuum at 250 °C for 20 h. In these samples, EDA is expected to deintercalate perfectly, taking into account the TG results. For reference, those of the host sample of 1T-TiSe$_2$ and as-intercalated samples are also shown. It is found that the post-annealed sample of (C$_2$H$_8$N$_2$)$_y$TiSe$_2$ returns to 1T-TiSe$_2$ from Phase II due to the deintercalation of EDA as expected. In fact, the $c$-axis length decreases from 10.107(4) Å to 6.0(6) Å, as listed in Table I.

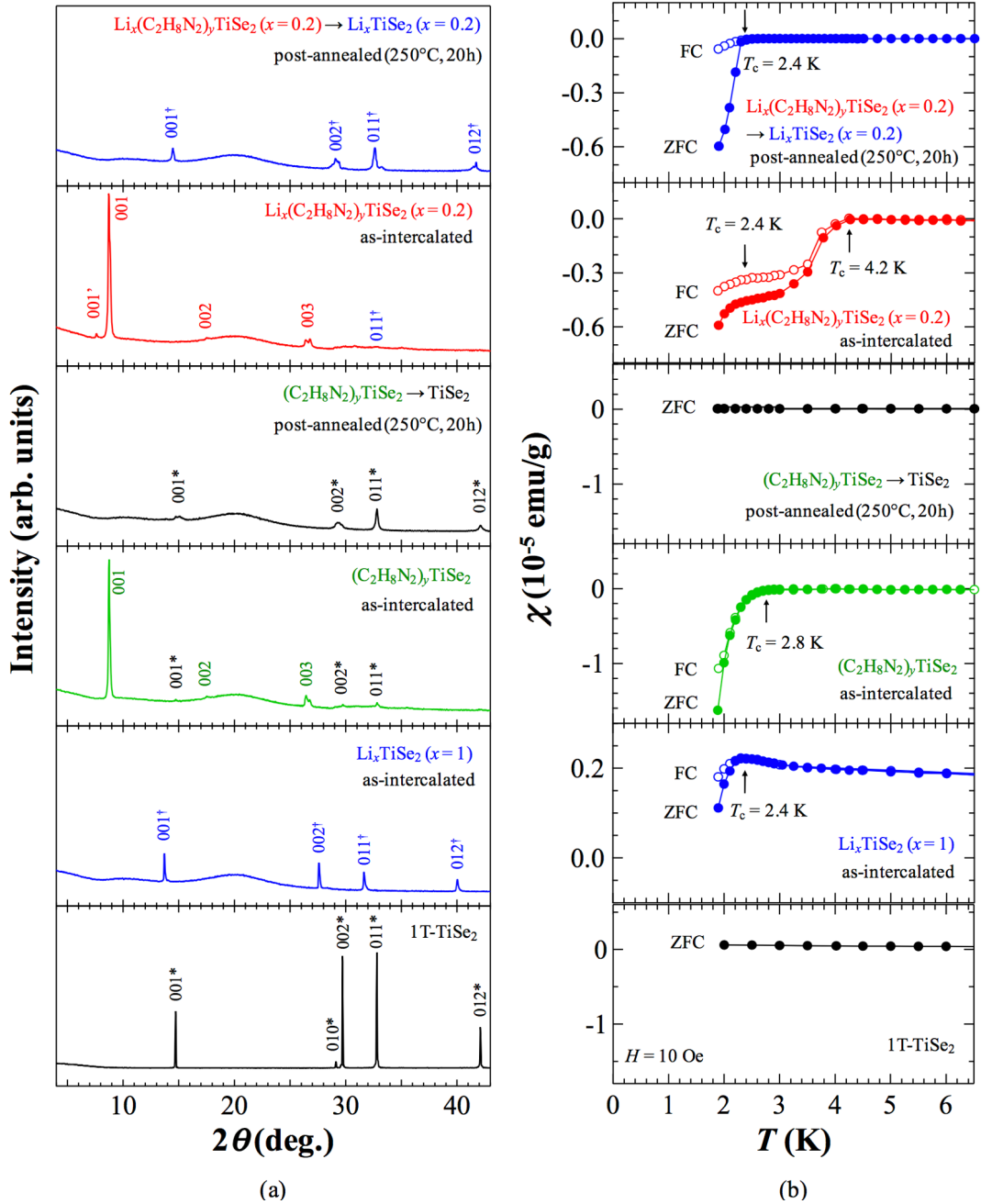

Fig. 7. (a) Powder x-ray diffraction patterns of $(C_2H_8N_2)_yTiSe_2$ and $Li_x(C_2H_8N_2)_yTiSe_2$ ($x = 0.2$) post-annealed in vacuum at 250°C for 20 h, using CuKα radiation. Those of the host sample of 1T-$TiSe_2$ and as-intercalated samples before the post-annealing are also shown for reference. Indices with an asterisk, a dagger, no mark and a quotation mark represent 1T-$TiSe_2$, Phase I, Phase II and Phase III of $Li_x(C_2H_8N_2)_yTiSe_2$. All indices are based on the 1T-type structure ($P\bar{3}m1$). The broad peak around $2\theta = 20°$ is due to the airtight sample-holder. (b) Temperature dependences of the magnetic susceptibility $\chi$ in a magnetic field of 10 Oe on zero-field cooling (ZFC) (closed circles) and on field cooling (FC) (open circles) for $(C_2H_8N_2)_yTiSe_2$ and $Li_x(C_2H_8N_2)_yTiSe_2$ ($x = 0.2$) post-annealed in vacuum at 250°C for 20 h. Those of 1T-$TiSe_2$ and as-intercalated samples are also shown for reference.

On the other hand, the post-annealed sample of $Li_x(C_2H_8N_2)_yTiSe_2$ ($x$ = 0.2) is found to return to Phase I of $Li_xTiSe_2$ ($x$ = 0.2) from Phase II and Phase III due to the deintercalation of EDA, because $Li^+$ ions are known not to deintercalate at a temperature as low as 250°C.[40] In fact, the $c$-axis length is 6.13(8) Å and longer than that of 1T-$TiSe_2$, as listed in Table I.

Figure 7(b) shows the temperature dependences of $\chi$ in a magnetic field of 10 Oe on ZFC and FC for post-annealed samples of $(C_2H_8N_2)_yTiSe_2$ and $Li_x(C_2H_8N_2)_yTiSe_2$ ($x$ = 0.2). For reference, those of the host sample of 1T-$TiSe_2$ and as-intercalated samples are also shown. It is found that the post-annealed sample of $(C_2H_8N_2)_yTiSe_2$ shows no superconductivity. This is reasonable, because the post-annealed sample has returned to 1T-$TiSe_2$ as mentioned above. On the other hand, the two-step superconducting transition at $T_c$ = 4.2 K and 2.4 K in the as-intercalated sample of $Li_x(C_2H_8N_2)_yTiSe_2$ ($x$ = 0.2) is found to change to one superconducting transition at 2.4 K through the post-annealing. The value of $T_c$ = 2.4 K is the same as that of $Li_xTiSe_2$ ($x$ = 1). This is also reasonable, because the post-annealed sample has returned to $Li_xTiSe_2$ ($x$ = 0.2), though the Li content is different. Accordingly, it is concluded that $T_c$ of Phase I in only Li-intercalated samples of $Li_xTiSe_2$ is 2.4 K and is not so dependent on $x$ probably due to the inhomogeneous distribution of $Li^+$ ions in a sample. This result strongly supports the conclusion that the co-intercalation of Li and EDA is necessary to enhance $T_c$ up to 4.2 K. This enhancement of $T_c$ through the co-intercalation is speculated to be owing to the multiplier effect of (1) the enhancement of the two-dimensionality of the crystal structure and/or the electronic structure due to the expansion of the interlayer spacing between $TiSe_2$ layers and (2) the enlargement of the electronic density of states at the Fermi level due to the charge transfer from intercalated Li to $TiSe_2$ layers.

Figure 8(a) shows the temperature dependence of $\rho$ for the $(C_2H_8N_2)_yTiSe_2$ sample post-annealed in vacuum at 250°C for 20 h. For reference, those of the host sample of 1T-$TiSe_2$ and the as-intercalated sample are also shown. It is found that, through the post-annealing, both the metallic behavior of $\rho$ and the superconducting transition disappear and that the CDW transition reappears instead. This is reasonable, because the as-intercalated sample of $(C_2H_8N_2)_yTiSe_2$ has returned to 1T-$TiSe_2$ through the post-annealing as mentioned above. Since EDA is regarded as deintercalating perfectly through the post-annealing, these results reconfirm that the expansion of the interlayer spacing between $TiSe_2$ layers due to the intercalation of EDA suppresses the CDW transition, leading to the appearance of superconductivity with $T_c$ = 2.8 K. $T_c$ of Phase II in both Li- and EDA-intercalated samples of $Li_x(C_2H_8N_2)_yTiSe_2$ is 2.4 – 2.8 K, as mentioned before. Therefore, $T_c$ of Phase II of $(C_2H_8N_2)_yTiSe_2$ seems not to be affected so much by the co-intercalation of Li.

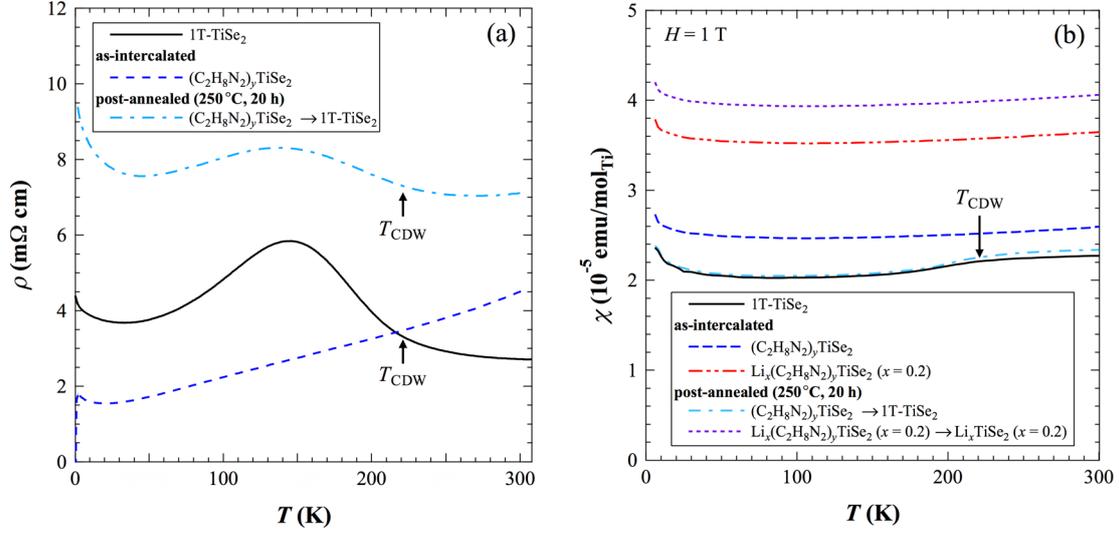

Fig. 8. (a) Temperature dependence of the electrical resistivity $\rho$ for $(C_2H_8N_2)_y$TiSe$_2$ post-annealed in vacuum at 250°C for 20 h, which were pelletized at room temperature without heat treatment. Those of the host sample of 1T-TiSe$_2$ and the as-intercalated sample before the post-annealing, which were pelletized at room temperature, are also shown for reference. (b) Temperature dependences of the magnetic susceptibility $\chi$ in a magnetic field of 1 T on zero-field cooling for $(C_2H_8N_2)_y$TiSe$_2$ and Li$_x$(C$_2$H$_8$N$_2$)$_y$TiSe$_2$ ($x$ = 0.2) post-annealed in vacuum at 250°C for 20 h. Those of 1T-TiSe$_2$ and as-intercalated samples are also shown for reference.

confirmed in the temperature dependences of $\chi$ in a magnetic field of 1 T on ZFC, as shown in Fig. 8(b), because $\chi$ of the post-annealed sample of $(C_2H_8N_2)_y$TiSe$_2$ is almost the same as that of 1T-TiSe$_2$. In Fig. 8(b), the change of $\chi$ through the post-annealing is shown for Li$_x$(C$_2$H$_8$N$_2$)$_y$TiSe$_2$ ($x$ = 0.2) also. It is found that $\chi$ increases through the deintercalation of EDA. This may be due to the possible transfer of valence electrons of intercalated Li, which were partly used for the bonding with EDA, to TiSe$_2$ layers, leading to the increase in the Pauli paramagnetism.

## 4. Conclusions

We have successfully synthesized new intercalation superconductors of Li$_x$(C$_2$H$_8$N$_2$)$_y$TiSe$_2$ and Li$_x$(C$_6$H$_{16}$N$_2$)$_y$TiSe$_2$ with $T_c$ = 4.2 K via the co-intercalation of lithium and EDA or HMDA into 1T-TiSe$_2$. Moreover, it has been found that Li$_x$TiSe$_2$ and $(C_2H_8N_2)_y$TiSe$_2$ also show superconductivity with $T_c$ = 2.4 K and 2.8 K, respectively.

($C_2H_8N_2$)$_y$TiSe$_2$ and Li$_x$($C_2H_8N_2$)$_y$TiSe$_2$ have been found to return to 1T-TiSe$_2$ and Li$_x$TiSe$_2$ through the deintercalation of EDA by the post-annealing, respectively.   These results indicate that both the electron doping due to the intercalation of Li and the expansion of the interlayer spacing between TiSe$_2$ layers due to the intercalation of diamines suppress the CDW transition, leading to the appearance superconductivity.   The result that the $T_c$ value of both Li- and EDA- or HMDA-intercalated Li$_x$($C_2H_8N_2$)$_y$TiSe$_2$ or Li$_x$($C_6H_{16}N_2$)$_y$TiSe$_2$ is higher than those of only Li-intercalated Li$_x$TiSe$_2$ and only EDA-intercalated ($C_2H_8N_2$)$_y$TiSe$_2$ is speculated to be owing to the multiplier effect of (1) the enhancement of the two-dimensionality of the crystal structure and/or the electronic structure due to the expansion of the interlayer spacing between TiSe$_2$ layers and (2) the enlargement of the electronic density of states at the Fermi level due to the charge transfer from intercalated Li to TiSe$_2$ layers.

**Acknowledgments**

This work was supported by JSPS KAKENHI (Grant Numbers 15K13512 and 16K05429).   One of the authors (K. S.) was supported by the Tohoku University Division for Interdisciplinary Advanced Research and Education.